\documentclass[showpacs,preprintnumbers,amsmath,amssymb]{revtex4}

\usepackage{graphicx}
\usepackage{epsfig,amsmath,amssymb}
\usepackage{hyperref}
\usepackage{epstopdf}
\usepackage{float}
\begin{document}

\title{Two novel immunization strategies for epidemic control in directed scale-free networks with nonlinear infectivity}

\author{Wei Shi, Junbo Jia, Pan Yang, Xinchu Fu}\altaffiliation{Corresponding author. Tel: +86-21-66132664; Fax: +86-21-66133292; E-mail: xcfu@shu.edu.cn}

\affiliation{\small Department of Mathematics, Shanghai University, Shanghai 200444, China}

\date{\today}

\begin{abstract}
\noindent In this paper, we propose two novel immunization strategies, i.e., combined immunization and
duplex immunization, for SIS model in directed scale-free networks, and obtain the epidemic thresholds
for them with linear and nonlinear infectivities. With the suggested two new strategies,
the epidemic thresholds after immunization are greatly increased. For duplex
immunization, we demonstrate that its performance is the best among all usual 
immunization schemes with respect to degree distribution. And for combined immunization scheme, 
we show that it is more effective than active immunization.
Besides, we give a comprehensive theoretical analysis on applying targeted immunization 
to directed networks. For targeted immunization strategy, we prove that 
immunizing nodes with large out-degrees are more effective than immunizing nodes 
with large in-degrees, and nodes with both large out-degrees and large in-degrees 
are more worthy to be immunized than
nodes with only large out-degrees or large in-degrees. Finally, some numerical analysis are 
performed to verify and complement our theoretical results. This work is the first to 
divide the whole population into different types and embed appropriate immunization 
scheme according to the characteristics of the population, and it will benefit
the study of immunization and control of infectious diseases on complex networks.

\vspace{0.2cm}


\end{abstract}


\maketitle


\section{Introduction}
\label{sec-1}

Devising effective immunization schemes is very important for the prevention and control of
infectious diseases and computer viruses. For implement of an immunization scheme, when a portion of
nodes (individuals) are immunized, those nodes can be thought as being
removed from the network and cannot infect or be infected by others, then the tolerance of
the network will be strengthened, so it is very important to choose key nodes to be immunized.
In the previous works, studies are mainly focused on immunization
schemes for susceptible-infected-susceptible (SIS) models~\cite{BogunaM2011,Moreno2002,Pastor-Satorras2012,Fu2008,Pastor2001,Tanimoto2011}. 
The epidemic threshold characterizes
the robustness of networks, and it signifies the critical value for a disease outbreak, when
infection rate of an epidemic is beyond it, the epidemic will spread on the network, and an epidemic vanishes
while the infection rate is below it.  To increase the threshold, designing effective immunization strategies
is very important~\cite{Immunization2002}. Up to now, many effective immunization schemes have been
proposed and studied, such as  acquaintance immunization~\cite{R2003}, random immunization~\cite{Pastor-Satorras2012}, targeted immunization~\cite{Pastor-Satorras2012},
active immunization~\cite{Fu2008}, greedy immunization~\cite{greedy2014}, dynamic
immunization~\cite{dynamic2015}, and so on.

Although many immunization strategies for epidemic models on complex networks have been studied extensively~\cite{Pastor2001,Immunization2002,bai2007,R2003,R2004,M2004},  most of them are based on
undirected networks~\cite{Pastor-Satorras2012,Fu2008}. Real-world networks are closely related to
directed networks, such as social networks, food webs, phone-call networks, the WWW~\cite{www2003}, etc.
The direction of nodes' edges plays an important role in the study of epidemic spread on
networks.  Diseases, viruses or information spread
out via their out-going edges and connect to others, while a susceptible node may be infected by its in-coming
edges. Therefore, studying immunization strategies in directed networks is more practical and
meaningful.

Many directed networks, such as the WWW~\cite{www2003} and social networks, have power-law degree
distributions of the form:
\begin{center}
$P(k)=C_{1}k^{-2-\gamma}$ and $Q(l)=C_{2}l^{-2-\gamma^{'}}$,
\end{center}
where $C_{1}$ and $C_{2}$ are normalization constants to guarantee $\sum_{k=m}^{M}P(k)=1$ and $\sum_{l=n}^{N}Q(l)=1$. Here we have $0<\gamma, \gamma^{'}\leq1$. Networks with power-law degree
distributions are called scale-free. Here we suppose that $m$ is the minimal out-degree, $M$
the maximal out-degree, $n$ the minimal in-degree, and $N$ the maximal in-degree
in the network.

In this paper, based on the heterogeneous mean-field theory~\cite{Pastor2001} and degree distribution, 
we mainly study four immunization strategies
for the SIS model on a directed network, including active immunization, targeted
immunization, combined immunization and duplex immunization. 
We obtain the epidemic thresholds for these
four immunization schemes, and compare effectiveness among them under the same immunization rate. 
Our results show that the proposed duplex immunization strategy is the most effective scheme
among all usual immunization schemes, including proportional immunization,  acquaintance
immunization, targeted immunization, active immunization, and the proposed combined immunization.
Besides, we divide the targeted immunization strategy into three cases in directed networks.
We prove that nodes with large out-degrees are more important than nodes with large in-degrees 
when targeted immunization is implemented. On the other hand, we demonstrate
that the nodes with both large in-degrees and large out-degrees are more 
worthy to be immunized than nodes with only large in-degrees or large out-degrees 
for targeted immunization scheme.
To illustrate and test the performance of the proposed immunization schemes, 
we present numerical simulations in directed BA network
in Figs~\ref{fig:1}-\ref{fig:3}, the numerical results are in accordance with our theoretical
results.

The rest of the paper is organized as follows. In Section~\ref{sec-2}, 
we establish an SIS model in a directed network and discuss epidemic 
thresholds with different infectivities. In Section~\ref{sec-3}, we first study
in detail the targeted immunization scheme in a directed network, then analyze 
the active immunization in a directed network. Besides, we propose two novel 
immunization strategies, and calculate the epidemic
thresholds for them, and compare their effectiveness with 
targeted immunization and active immunization.
In Section~\ref{sec-4}, we present numerical simulations. 
Finally, in Section~\ref{sec-5}, we conclude the paper.

\section{The SIS Model in directed networks}
\label{sec-2}

In this section, we investigate the SIS model on a directed network. 
Nodes of the directed network are divided into two groups: Susceptible
and Infected. Hereafter, we will denote a susceptible node by an S-node etc., for short. An S-node becomes
infected at rate $\nu$ if it contacts with an infected individual, and an I-node may recover and become
an S-node with probability $\delta$. Previous works have defined an effective spreading rate $\lambda=\frac{\nu}{\delta}$,  where we take a unit recovery rate $\delta=1$.  Let us denote the
densities of S- and I-nodes with in-degrees $k$ and out-degrees $l$ at
time $t$ by $s_{k,l}(t)$, $\rho_{k,l}(t)$, respectively, so we have
\begin{center}
  $s_{k,l}(t)+\rho_{k,l}(t)=1,$
\end{center}
 where follows the joint probability distribution $p(k, l)$. Then, the respective marginal
 probability distribution of the out-degrees and in-degrees reads as
\begin{center}
  $P(k)=\underset{l}{\sum}p(k,l),~~ Q(l)=\underset{k}{\sum}p(k,l),$
\end{center}
and their average degrees are
\begin{center}
 $ \langle k\rangle=\underset{k,l}{\sum}kp(k,l)=\underset{k}{\sum}kP(k), ~~\langle l\rangle=\underset{k,l}{\sum}lp(k,l)=\underset{l}{\sum}lQ(l).$
\end{center}
Then the SIS model can be written as the following ordinary differential equations:
\begin{equation}
\frac{d\rho_{k,l}}{dt}=\lambda k(1-\rho_{k,l}(t))\Theta(t)-\rho_{k,l}(t).  \label{1}
\end{equation}
Here we suppose that the connectivity of nodes is uncorrelated, then the probability of a randomly
selected outgoing link emanating form I-nodes at time $t$ is given by
\begin{equation}
\Theta(t)=\frac{\underset{k,l}{\sum}\varphi(k,l)p(k,l)\rho_{k,l}(t)}{\underset{k,l}{\sum}lp(k,l)}
=\frac{\underset{k,l}{\sum}\varphi(k,l)p(k,l)\rho_{k,l}(t)}{{\langle l\rangle}},\label{2}
\end{equation}
where $\varphi(k,l)$ denotes the infectivity of a node with degrees $(k,l)$.

Now, we  calculate the epidemic threshold for model (\ref{1}).
At the steady state, we have $\frac{d\rho_{k,l}}{dt}=0$ for 
all $k$ and $l$, from (\ref{1}) we have
\begin{equation}
\rho_{k,l}=\frac{\lambda k\Theta}{1+\lambda k\Theta},\notag
\end{equation}
substituting the above equation into (\ref{2}) we obtain a self-consistency equation for $\Theta$ as
follows:
\begin{equation}
\Theta=\frac{1}{\langle l\rangle}\underset{k,l}{\sum}\frac{\lambda\varphi(k,l)p(k,l)k\Theta}{1+\lambda k\Theta}\equiv f_{1}(\Theta).\notag
\end{equation}

If this equation has a solution $\Theta>0$ other than $\Theta=0$, then it corresponds to an endemic state.
Note that
\begin{eqnarray*}
f_{1}(1)&=&\frac{1}{\langle l\rangle}\underset{k,l}{\sum}\frac{\lambda\varphi(k,l)p(k,l)k}{1+\lambda k}<\frac{1}{\langle l\rangle}\underset{k,l}{\sum}lp(k,l)=1,\\
f'_{1}(\Theta)&=&\frac{1}{\langle l\rangle}\underset{k,l}{\sum}\varphi(k,l)p(k,l)\frac{\lambda k}{(1+\lambda k\Theta)^{2}}>0,\\
f''_{1}(\Theta)&=&\frac{1}{\langle l\rangle}\underset{k,l}{\sum}\varphi(k,l)p(k,l)\frac{-2(\lambda k(1-\delta(k,l)))^{2}}{(1+\lambda k\Theta)^{3}}<0,
\end{eqnarray*}
therefore, a nontrivial solution exists only if
\begin{equation}
\frac{df_{1}(\Theta)}{d\Theta}\bigg|_{\Theta=0}>1,\label{3}
\end{equation}
so we obtain the value of $\lambda$ yielding the inequality~(\ref{3}) which defines the critical epidemic threshold $\lambda_{c}$:
\begin{equation}
\lambda_c=\frac{\langle l\rangle}{\langle \varphi(k,l)k\rangle}, \label{4}
\end{equation}
where the $\lambda_c$ is a critical value for the infection rate $\lambda$: If $\lambda >\lambda_{c}$,
the disease will break out and persist on this network; Otherwise, when $\lambda< \lambda_{c}$, the disease will
gradually peter out. Hence, it is very crucial to increase $\lambda_{c}$ on
the network to prevent epidemic outbreak. We will give detailed analysis on this in Section~\ref{sec-3}.

From the equality~(\ref{4}), we can see that the infectivity $\varphi(k,l)$ also affects the value 
of the threshold $\lambda_c$,
so $\varphi(k,l)$ is also important for controlling the disease. For this
reason, we give further study of the infectivity in the following subsection.


\subsection{The epidemic threshold for the SIS model with nonlinear infectivity}
\label{sec-2-1}

For the SIS model on undirected scale-free networks~\cite{Fu2008}, $\varphi(k)$ indicates the infectivity
of a node with degree $k$. Previously, it was assumed that the larger the node degree, the
larger the value of $\varphi(k)$, and in~\cite{Pastor2001,Pastor2003,satorras2003,Pastor2002},
the $\varphi(k)$ is just equal to the node degree, that is, $\varphi(k)=k$,
in this case, the epidemic threshold $\lambda_{c}=0$
when networks' size is sufficiently large. However, in~\cite{bai2007,yang2007,z2006}, the authors
pointed out that large node with large $\varphi(k)$ is not always appropriate,
so they assumed that $\varphi(k)=A$, where $A$ is a constant, and they obtained a different epidemic
threshold $\lambda_{c}=\frac{1}{A}$, which is always positive. On the basis of this, in~\cite{z2009}
authors proposed a new nonlinear infectivity
$\varphi(k)=\frac{ak^{\alpha}}{1+bk^{\alpha}}$, and analysis its threshold on finite and infinite
networks.

Here in a directed scale-free network, we think both out-degrees and in-degrees play an important role
in infectivity $\varphi(k,l)$. At the early stage of a disease transmission, a susceptible individual may
get infected through out-going edges of infected individuals (in-coming edges of itself), then the
disease spreads out of its out-going edges and connects to other susceptible nodes.
When a susceptible individual has no in-coming edges, it cannot be infected by infected individuals;
similarly, it will not infect other susceptible individuals without out-going edges even if it was infected.

Base on the analysis above, we give a nonlinear infectivity $\varphi(k,l)$ in a directed scale-free
network as follows:
\begin{equation}
\varphi(k,l)=\frac{al^{\alpha}}{1+bl^{\alpha}}\cdot\frac{ck^{\beta}}{1+dk^{\beta}},\label{5}
\end{equation}
where $0\leq\alpha, \beta\leq1, a>0$ and $b, c, d\geq0$. In Eq.~(\ref{5}), when $k$ (or $l$) is very small,
we can simply regard $\frac{ck^{\beta}}{1+dk^{\beta}}$ (or $\frac{al^{\alpha}}{1+bl^{\alpha}}$)
as  $0$, which means this node cannot infect others (or cannot be infected by others); and during the
epidemic spreading process, the disease spreads out of infected individuals' out-going edges, so the in-degrees
of susceptible nodes is relatively more important than the out-degrees at the early stage, based on
this, we choose $a>0$ rather than $a=0$;
and here we divide the $\varphi(k,l)$ into four main cases:

(1)$\varphi(k,l)=\frac{ac}{(1+b)(1+d)}$ when $\alpha=0, \beta=0$, which means infectivity is a
constant;

(2)$\varphi(k,l)=al$ when $\alpha=1, b, c=0$;

(3)$\varphi(k,l)=al^{\alpha}$ when $0<\alpha<1, b, c=0$;

(4)if $b, c\neq 0$, then $\varphi(k,l)=\frac{al^{\alpha}}{1+bl^{\alpha}}\cdot\frac{ck^{\beta}}{1+dk^{\beta}}$, and it becomes gradually saturated with the increasing of in-degree $k$ and out-degree $l$. Finally,
it will converges to a constant $\varphi(k,l)=\frac{ac}{bd}$.

Substituting case~(1) and case~(2) into Eq.~(\ref{4}), we obtain two different epidemic thresholds
as follows: $\lambda_{c}^{1}=\frac{ac \langle l\rangle}{(1+b)(1+c)\langle k\rangle}$
and $\lambda_{c}^{2}=\frac{1}{a\langle k\rangle}$, which were partially studied in~\cite{w2012}; and with
sufficiently large $k$ and $l$, $\lambda_{c}^{1}=\frac{\langle ac\rangle}{\langle(1+b)(1+c)\rangle}$
and $\lambda_{c}^{2}=0$.

When $\varphi(k,l)=al^{\alpha}$ in case~(3), we have $\langle \varphi(k,l)k\rangle=a\sum_{k}kP(k)\sum_{l}l^{\alpha}Q(l)$. By using a continuous approximation,
we obtain
\begin{equation}
\langle k\varphi(k,l)\rangle=a\int_{m}^{\infty}\int_{m}^{\infty}k\cdot k^{-2-\gamma}\cdot l^{\alpha}\cdot l^{-2-\gamma'}dkdl=a\int_{m}^{\infty}\frac{1}{k^{1+\gamma}}dk\cdot\int_{m}^{\infty}\frac{1}{l^{1+\gamma'-\alpha}}dl.\label{6}
\end{equation}
From the above equation, we can conclude that the (\ref{6}) is bounded when $\alpha<\gamma'$.
As a result, the epidemic threshold is
\begin{equation}
\lambda_{c}^{3}=\frac{m^{\gamma-\alpha}\gamma(1-\frac{\alpha}{\gamma'})}{a},\label{7}
\end{equation}
which is always positive regardless of the size of out-degrees and in-degrees. We believe this is a interesting result, which is different from the result of a vanished threshold $\lambda_c=0$ given
in~\cite{Tanimoto2011}.

When $\varphi(k,l)=\frac{al^{\alpha}}{1+bl^{\alpha}}\cdot\frac{ck^{\beta}}{1+dk^{\beta}}$ ,
where we have $0<\alpha, \beta\leq1$ and $b, d\neq0$ in case(4), then
\begin{equation}
\langle k\varphi(k,l)\rangle=ac\sum_{k,l}p(k,l)k\frac{al^{\alpha}}{1+bl^{\alpha}}\cdot\frac{ck^{\beta}}{1+dk^{\beta}}
=\int_{n}^{\infty}Q(l)\frac{al^{\alpha}}{1+bl^{\alpha}}\cdot\int_{m}^{\infty}P(k)\frac{ck^{\beta}}{1+dk^{\beta}},
\end{equation}
similar to the above analysis in case~(3), we can find that $\langle k\varphi(k,l)\rangle$ is always
bounded, then we obtain that the threshold $\lambda_c^{4}$ is always a positive value.

Through analysis and calculation in this subsection, we obtain the different 
epidemic thresholds $\lambda_c^i~(i=1,2,3,4)$ 
for the four cases, and compare them with previous results. In Figs.~\ref{fig:1}-\ref{fig:3}, we present
numerical analysis for different infectivities $\varphi(k,l)$, it clearly shows that the
different infectivity $\varphi(k,l)$'s value result in different thresholds; and with nodes' infectivity
grows, $\lambda_c$ become smaller and smaller, so the network robustness against epidemics become
weaker and weaker.


\section{The SIS model with immunization}
\label{sec-3}

From the analysis in Section~\ref{sec-2}, we know that higher threshold $\lambda_{c}$ indicates
better robustness against the outbreak of an epidemic on a network. Hence, designing an appropriate
immunization strategy is important for effectively controlling the spread of the
epidemic. And the SIS model is known as a more appropriate model than SIR model to
study immunization schemes at the early time of epidemic transmission because the effects and recovery
and death can be ignored, and this is the optimal time to apply immunization strategies in order
to prevent and control epidemic outbreaks. In this section, we study the SIS model with various immunization
strategies and compare their effectiveness among them for the same average immunization rate.

\subsection{Targeted immunization in directed networks}
\label{sec-3-1}

The targeted immunization~\cite{P2002} is known as the best strategy on heterogeneous networks,
but we still lack a comprehensive understanding when applying it to directed networks.
The traditional targeted immunization~\cite{Pastor-Satorras2012,P2002} on undirected networks
is to pick up the nodes with connectivity $k>\kappa$ to immunize, such as Eq.(14)
in~\cite{Fu2008}. In~\cite{w2012}, Wang first studied the SIS model with targeted immunization
in directed networks, but only immunize nodes with large out-degrees.

We realize that in the real-life systems with targeted immunization scheme, only select the large
out-degree's nodes to immunize~\cite{w2012} may not always be appropriate. Beyond that, the in-degrees
may also play a significant role in the epidemic immunization process; as we discussed in Section~\ref{sec-2}, even the out-degrees of some nodes is very high, but those nodes may not always be infective with
their in-degrees are too small. Otherwise, the nodes of high in-degrees with low out-degrees may
not be infective as well.

So here we divide targeted immunization schemes into three cases to 
further compare their effectiveness: $(1)$ Immunize the nodes 
with $k>\pi_{1}$; $(2)$ Immunize the nodes with $l>\pi_{2}$;
and $(3)$ Immunize the nodes with $k >\pi_{3}$ and $l>\pi_{4}$. Next in this subsection, we
give a deep theoretical analysis on target immunization in directed networks under these three
conditions, and to find an optimal one beyond them. Here we considere $\varphi(k,l)=A$ as a
positive constant,  we define the immunization rate $\delta_{k,l}^{i}(0<\delta_{k,l}^{i}\le 1)$ by
\begin{equation}
  \delta_{k,l}^{1}=\begin{cases}
1, ~k>\pi_{1}\\
a,~ k=\pi_{1}~,\\
0, ~k<\pi_{1}
\end{cases}
\delta_{k,l}^{2}=\begin{cases}
1, ~l>\pi_{2}\\
b, ~l=\pi_{2}~,\\
0, ~l<\pi_{2}
\end{cases}
\delta_{k,l}^{3}=\begin{cases}
1,~k>\pi_{3}~~\mbox{and}~~l>\pi_{4}\\
c, ~k=\pi_{3}~~\mbox{and}~~l=\pi_{4}~,\\
0, ~\mbox{otherwise}
\end{cases}\label{13}
\end{equation}
where $0<a,b,c\leq1$, and $\underset{k,l}{\sum}\delta_{k,l}^{i}p(k,l)=\langle \delta_{k,l}^{i}\rangle$ 
$(i=1, 2, 3),$ and $\langle \delta_{k,l}^{i}\rangle$ are the average immunization rates. Then the epidemic
dynamics model (\ref{1}) becomes
\begin{equation}
\frac{d\rho_{k,l}}{dt}=\lambda k(1-\delta_{k,l})(1-\rho_{k,l}(t))\Theta(t)-\rho_{k,l}(t).\label{8}
\end{equation}

At the steady state,  we have the condition $\frac{d\rho_{k,l}}{dt}=0$ for all $k$ and $l$. So we can get from (\ref{8}) that
\begin{equation}
\rho_{k,l}=\frac{\lambda k(1-\delta_{k,l})\Theta}{1+\lambda k(1-\delta_{k,l})\Theta}.\notag
\end{equation}

Substituting these into Eq.~\ref{2}, we obtain a self-consistency equation for $\Theta$
as follows:
\begin{equation}
\Theta=\frac{1}{\langle l\rangle}\underset{k,l}{\sum}\frac{\lambda\varphi(k,l)p(k,l)k(1-\delta_{k,l})\Theta}{1+\lambda k(1-\delta_{k,l})\Theta}\equiv f_{2}(\Theta),\notag
\end{equation}
therefore, we can obtain the threshold for model~(\ref{8}):
\begin{equation}
\hat{\lambda}_c=\frac{\langle l\rangle}{\langle\varphi(k,l)k\rangle-\langle\varphi(k,l)k\delta_{k,l}\rangle}.\label{10}
\end{equation}

Substituting (\ref{13}) into (\ref{10}) we  obtain three epidemic thresholds
with targeted immunization~(1)~(TGA), targeted immunization~(2)~(TGB) and targeted
immunization~(3)~(TGC), respectively:
\begin{equation}
\hat{\lambda}_{c}^{1}=\frac{\langle l\rangle}{\langle\varphi(k,l)k\rangle-\langle\varphi(k,l)k\delta_{k,l}^{1}\rangle}=\frac{\langle l\rangle}{A(\langle k\rangle-\langle k\delta_{k,l}^{1}\rangle)},\label{14}
\end{equation}
\begin{equation}
\hat{\lambda}_{c}^{2}=\frac{\langle l\rangle}{\langle\varphi(k,l)k\rangle-\langle\varphi(k,l)k\delta_{k,l}^{2}\rangle}=\frac{\langle l\rangle}{A(\langle k\rangle-\langle k\delta_{k,l}^{2}\rangle)},\label{15}
\end{equation}
\begin{equation}
\hat{\lambda}_{c}^{3}=\frac{\langle l\rangle}{\langle\varphi(k,l)k\rangle-\langle\varphi(k,l)k\delta_{k,l}^{3}\rangle}=\frac{\langle l\rangle}{A(\langle k\rangle-\langle k\delta_{k,l}^{3}\rangle)}.\label{16}
\end{equation}

Higher epidemic thresholds indicates better performance of immunization schemes, through three epidemic
thresholds (\ref{14})$-$(\ref{16}), now we discuss the effectiveness of different targeted
immunization schemes by comparing these three values:
$\langle k\delta_{k,l}^{1}\rangle$, $\langle k\delta_{k,l}^{2}\rangle$ and $\langle k\delta_{k,l}^{3}\rangle$, the bigger one corresponds to better effectiveness.

Intuitively, we think the TGA is more effective than TGB, and the TGC is the optimal
targeted immunization scheme in directed networks. During the immunization process, the
in-coming links of the immunized S-nodes comes from the I-nodes and S-nodes itself, therefore,
when we implement a target immunization on S-nodes, this in-coming links with $l>\pi_{2}$
comes from S-nodes itself which are harmless but are immunized at the same time, so
the TGB may less effective than TGA.

In addition, as we explained in previous sections, the in-degrees also play a significant role in the
immunization process, so it would be better to immunize the nodes with both large in-degrees
and large out-degrees, which will be further discussed in this subsection.

Note that
\begin{eqnarray*}
\langle k\delta_{k,l}^{1}\rangle=\underset{k,l}{\sum}k\delta_{k,l}^{1}p(k,l)=\overset{M}{\underset{k=\pi_{1}}{\sum}}kP(k),~~~~~~~~~~\\
\langle k\delta_{k,l}^{2}\rangle=\underset{k,l}{\sum}k\delta_{k,l}^{2}p(k,l)=\overset{N}{\underset{l=\pi_{2}}{\sum}}Q(l)\overset{M}{\underset{k=m}{\sum}}kP(k),\\
\langle k\delta_{k,l}^{3}\rangle=\underset{k,l}{\sum}k\delta_{k,l}^{3}p(k,l)=\overset{N}{\underset{l=\pi_{4}}{\sum}}Q(l)\overset{M}{\underset{k=\pi_{3}}{\sum}}kP(k).
\end{eqnarray*}
Under the same average immunization rate, which we take the average, then
\begin{align}
\delta_{k,l}^{1}=\delta_{k,l}^{2}=\delta_{k,l}^{3}
 &=\langle \delta_{k,l}^{1}\rangle= \langle\delta_{k,l}^{2}\rangle=\langle\delta_{k,l}^{3}\rangle\\ \notag
 &=\overset{M}{\underset{k=\pi_{1}}{\sum}}P(k)=\overset{N}{\underset{l=\pi_{2}}{\sum}}Q(l)=\overset{M}{\underset{k=\pi_{3}}{\sum}}\overset{N}{\underset{l=\pi_{4}}{\sum}}P(k)Q(l),
 \end{align}
through the above equations, we have
\begin{align}
\langle k\delta_{k,l}^{1}\rangle-\langle k\delta_{k,l}^{2}\rangle\notag
&=\overset{M}{\underset{k=\pi_{1}}{\sum}}kP(k)-\overset{N}{\underset{l=\pi_{2}}{\sum}}Q(l)\overset{M}{\underset{k=m}{\sum}}kP(k)\\ \notag
&=\overset{M}{\underset{k=\pi_{1}}{\sum}}kP(k)-\big(\overset{\pi_{1}-1}{\underset{k=m}{\sum}}kP(k)+ \overset{M}{\underset{k=\pi_{1}}{\sum}}kP(k)  \big)\overset{M}{\underset{k=\pi_{1}}{\sum}}P(k)\\ \notag
&=\overset{M}{\underset{k=\pi_{1}}{\sum}}kP(k)\overset{\pi_{1}-1}{\underset{k=m}{\sum}}P(k)-\overset{\pi_{1}-1}{\underset{k=m}{\sum}}kP(k)\overset{M}{\underset{k=\pi_{1}}{\sum}}P(k)\\ \notag
&=\overset{M}{\underset{\overline{k}=\pi_{1}}{\sum}}\overset{\pi_{1}-1}{\underset{k=m}{\sum}}\overline{k}P(\overline{k})P(k)-\overset{M}{\underset{\overline{k}=\pi_{1}}{\sum}}\overset{\pi_{1}-1}{\underset{k=m}{\sum}}kP(k)P(\overline{k})\\  \notag
&=\overset{M}{\underset{\overline{k}=\pi_{1}}{\sum}}\overset{\pi_{1}-1}{\underset{k=m}{\sum}}P(\overline{k})P(k)\big(\overline{k}-k\big)\\ \notag
&>0,\notag
 \end{align}
so it proves that the effectiveness of TGA is better than TGB.

Next, we discuss the effectiveness between TGC and TGA,TGB. Here, for better comparison,
we set $\pi_{1}=\langle k\rangle$ , $\pi_{2}=\langle l\rangle$, and we find when $\pi_{3}>\langle k\rangle$
and $\pi_{4}>\xi$, where $\xi$ is a positive constant (can fetch its value as $\langle l\rangle$),
the TGC is better than TGA and TGB.

Note that
\begin{align}
\langle k\delta_{k,l}^{3}\rangle-\langle k\delta_{k,l}^{2}\rangle \notag
&=\overset{N}{\underset{l=\pi_{4}}{\sum}}Q(l)\overset{M}{\underset{k=\pi_{3}}{\sum}}kP(k)
-\overset{N}{\underset{l=\pi_{2}}{\sum}}Q(l)\overset{M}{\underset{k=m}{\sum}}kP(k)  \\ \notag
&=\overset{M}{\underset{k=\pi_{3}}{\sum}}\overset{N}{\underset{l=\pi_{4}}{\sum}}Q(l)kP(k)-
\overset{M}{\underset{k=\pi_{3}}{\sum}}\overset{N}{\underset{l=\pi_{4}}{\sum}}Q(l)P(k)\overset{M}{\underset{k=m}{\sum}}kP(k)\\ \notag
&=\overset{M}{\underset{k=\pi_{3}}{\sum}}\overset{N}{\underset{l=\pi_{4}}{\sum}}Q(l)P(k)\big(k-\overset{M}{\underset{k=m}{\sum}}kP(k)\big) \\ \notag
&=\overset{M}{\underset{k=\pi_{3}>\langle k\rangle}{\sum}}\overset{N}{\underset{l=\pi_{4}}{\sum}}Q(l)P(k)\big(k-\langle k\rangle\big) \\ \notag
&>0 ;\notag
 \end{align}
\begin{align}
\langle k\delta_{k,l}^{3}\rangle-\langle k\delta_{k,l}^{1}\rangle \notag
&=\overset{N}{\underset{l=\pi_{4}}{\sum}}Q(l)\overset{M}{\underset{k=\pi_{3}}{\sum}}kP(k)-\overset{M}{\underset{k=\pi_{1}}{\sum}}kP(k)~~~~~~~~~ \\ \notag
&=\overset{M}{\underset{k=\pi_{3}}{\sum}}kP(k)\frac{\overset{M}{\underset{k=\pi_{1}}{\sum}}P(k)}
{\overset{M}{\underset{k=\pi_{3}}{\sum}}P(k)}-\overset{M}{\underset{k=\pi_{1}}{\sum}}kP(k)~~~~~~~~~ \\ \notag
&=\frac{\overset{M}{\underset{k=\pi_{3}}{\sum}}kP(k)\overset{M}{\underset{\widehat{k}=\pi_{1}}{\sum}}P(\widehat{k})
-\overset{M}{\underset{k=\pi_{3}}{\sum}}P(k)\overset{M}{\underset{\widehat{k}=\pi_{1}}
{\sum}}\widehat{k}P(\widehat{k})}{\overset{M}{\underset{k=\pi_{3}}{\sum}}P(k)} \\ \notag
&=\frac{\overset{M}{\underset{k=\pi_{3}>\langle k\rangle}{\sum}}\overset{M}{\underset{\widehat{k}=\pi_{1}=\langle k\rangle}{\sum}}P(k)P(\widehat{k})\big(k-\widehat{k}\big)}{\overset{M}{\underset{k=\pi_{3}}{\sum}}P(k)}\\ \notag
&>0,
 \end{align}
therefore, the epidemic thresholds of TGC is greater than TGA's and TGB's, this means
that the performance of TGC is better than TGA and TGB. Either the infectivity is linear or nonlinear,
this conclusion is always valid. Figs.~\ref{fig:2}(c)-(d) and Figs.~\ref{fig:3}(c)-(d) below
show this in details.

In this section, we divide the classic targeted immunization scheme into three cases in 
directed networks, the discussion and comparison of those three cases are carried out in depth.
Now we have given the analytical comparison of the three different epidemic
thresholds (see (\ref{14})$-$(\ref{16})). We prove that the nodes with both large in-degrees and
large out-degrees are more worthy to be immunized during target immunization process in
directed networks. Besides, immunizing nodes with large out-degrees are more efficient than immunizing nodes with large in-degrees for targeted immunization scheme.

\subsection{Active immunization in directed networks }
\label{sec-3-2}

The classic active immunization~\cite{Fu2008} is to immunize its neighbors with degrees $k\geq\kappa$
on undirected scale-free networks, here we will generalize it to directed networks and calculate
its epidemic threshold. Then, the epidemic dynamics model becomes
\begin{equation}
\frac{d\rho_{k,l}}{dt}=\lambda k(1-\rho_{k,l}(t))\Theta(t)-(1+\overline{\delta}_{k,l})\rho_{k,l}(t)\label{17},
\end{equation}
where
\begin{center}
  $\overline{\delta}_{k,l}=\underset{k}{\sum}\frac{kP(k)}{\langle k \rangle}\delta_{k,l}^{1},$
\end{center}
and $\delta_{k,l}^{1}$ is defined in (\ref{13}).

By adopting $\frac{d\rho_{k,l}}{dt}=0$  leads to
\begin{equation}
\Theta=\frac{\lambda\Theta}{\langle l\rangle}\sum_{k,l}\frac{k\varphi(k,l)p(k,l)}{\lambda k\Theta+1+\overline{\delta}_{k,l}}\equiv f_{3}(\Theta);\notag
\end{equation}
therefore, we obtain the epidemic threshold
\begin{equation}
\tilde{\lambda}_{c}=\frac{\langle l\rangle}{\sum_{k,l}(1+\overline{\delta}_{k,l})^{-1}k\varphi(k,l)p(k,l)}.\notag
\end{equation}
Note that
\begin{equation}
\overline{\delta}_{k,l}=\sum_{k}\frac{kP(k)}{\langle k\rangle}\delta_{k,l}^{1}=\frac{\langle k\delta_{k,l}^{1}\rangle}{\langle k\rangle},\notag
\end{equation}
we have
\begin{equation}
\tilde{\lambda}_{c}=\frac{\langle kl\rangle+\langle l\rangle\langle k\delta_{k,l}^{1}\rangle}{\langle k^{2}\varphi(k,l)\rangle}.\label{18}
\end{equation}
Compare (\ref{10}) with (\ref{18}), we have
\begin{align}
  \hat{\lambda}_{c}-\tilde{\lambda}_{c}&=\frac{\langle l\rangle}{\langle\varphi(k,l)k\rangle-\langle\varphi(k,l)k\delta_{k,l}\rangle}-\frac{\langle kl\rangle+\langle l\rangle\langle k\delta_{k,l}^{1}\rangle}{\langle k^{2}\varphi(k,l)\rangle}\notag\\
  &=\frac{\langle kl\rangle\langle\varphi(k,l)k\delta_{k,l}\rangle-\langle kl\varphi(k,l)\rangle\langle k\delta_{k,l}^{1}\rangle+\langle l\rangle\langle k\delta_{k,l}^{1}\rangle\langle\varphi(k,l)k\delta_{k,l}\rangle}{\big(\langle k^{3}\varphi^{2}(k,l)\rangle-\langle\varphi(k,l)k\delta_{k,l}\rangle\langle k^{2}\varphi(k,l)\rangle\big)}.\label{19}
\end{align}
To better compare the target immunization and active immunization,
we set the same average immunization rate
\begin{equation}
\overline{\delta}_{k,l}=\delta_{k,l},\notag
\end{equation}
therefore,
\begin{equation}
\frac{\langle k\delta_{k,l}^{1}\rangle}{\langle k\rangle}=\langle\delta_{k,l}\rangle.\label{20}
\end{equation}
From  (\ref{19}) and (\ref{20}), we have
\begin{align}
\hat{\lambda}_c-\tilde{\lambda}_{c}=\frac{\langle kl\rangle\langle\varphi(k,l)k\delta_{k,l}\rangle-\langle kl\varphi(k,l)\rangle\langle k\rangle\langle\delta_{k,l}\rangle+B}{A},\label{21}
\end{align}
where $A=\langle k^{3}\varphi^{2}(k,l)\rangle-\langle\varphi(k,l)k\delta_{k,l}\rangle\langle k^{2}\varphi(k,l)\rangle$,  and  $B=\langle l\rangle\langle k\delta_{k,l}^{a}\rangle\langle\varphi(k,l)k\delta_{k,l}\rangle$.

Note that $\langle kl\varphi(k,l)\rangle\langle k\rangle\langle\delta_{k,l}\rangle=\langle kl\rangle\langle\varphi(k,l)k\delta_{k,l}\rangle+\sigma'$. There may be appropriate
small $\pi_{3}$, $\pi_{4}$, such that $\sigma'$ is relatively smaller than $B$. So we can
obtain that $\hat{\lambda}_c>\tilde{\lambda}_{c}$,
which means that the target immunization scheme is more effective than active immunization under
the same average immunization rate, Fig.~\ref{fig:1}(a) below illustrates this
conclusion.

\subsection{Combined immunization}
\label{sec-3-3}

In this section we propose a new immunization scheme, combined immunization: 
Choose a susceptible node and immunize
its neighbors whose in-degrees $l>\kappa_{1}$, and choose an infected node to immunize its
neighbors whose out-degrees $k>\kappa_{2}$ at the same time. Then the epidemic
dynamics model becomes:
\begin{equation}
\frac{d\rho_{k,l}}{dt}=\lambda k(1-\rho_{k,l}(t))(1-\overline{\delta}_{l})\Theta(t)-(1+\overline{\delta}_{k})\rho_{k,l}(t),\label{22}
\end{equation}
where
\begin{equation}
  \overline{\delta}_{l}=\underset{l}{\sum}\frac{lQ(l)}{\langle l \rangle}\delta_{l}, \overline{\delta}_{k}=\underset{k}{\sum}\frac{kP(k)}{\langle k \rangle}\delta_{k},\notag
\end{equation}

\begin{center}
    $\delta_{l}=\begin{cases}
1,~ l>\kappa_{1}\\
d,~ l=\kappa_{1}~,\\
0,~ l<\kappa_{1}
\end{cases}$
 $\delta_{k}=\begin{cases}
1, ~k>\kappa_{2}\\
e, ~k=\kappa_{2}~,\\
0, ~k<\kappa_{2}
\end{cases}$
\end{center}
and $0<d,e\leq1$.

In the early stages of disease transmission,  there may be quite a lot of susceptible
individuals and infected individuals; therefore, to immunize them at the same time may
be more appropriate. We show this rigorously below.

By letting $\frac{d\rho_{k,l}}{dt}=0$, than
substitute it into (\ref{2}), model (\ref{22}) leads to
\begin{equation}
\Theta=\frac{\lambda\Theta}{\langle l\rangle}\underset{k,l}{\sum}\frac{k\varphi(k,l)p(k,l)(1-\overline{\delta}_{l})}{\lambda k(1-\overline{\delta}_{l})\Theta+1+\overline{\delta}_{k}}\equiv f_{4}(\Theta).\notag
\end{equation}

So the epidemic threshold for model (\ref{22}) is
\begin{equation}
\bar{\lambda}_{c}=\frac{\langle l\rangle}{\underset{k,l}{\sum}(1+\overline{\delta}_{k})^{-1}k\varphi(k,l)p(k,l)(1-\overline{\delta}_{l})}.\notag
\end{equation}

Due to
\begin{center}
  $\overline{\delta}_{l}=\underset{l}{\sum}\frac{lQ(l)}{\langle l \rangle}\delta_{l}=\frac{\langle l\delta_{l}\rangle}{\langle l\rangle}$,
  $\overline{\delta}_{k}=\underset{k}{\sum}\frac{kP(k)}{\langle k \rangle}\delta_{k}=\frac{\langle k\delta_{k}\rangle}{\langle k\rangle}$.
\end{center}

We obtain that
\begin{equation}
\bar{\lambda}_{c}=\frac{\langle l^{2}\rangle(\langle k\rangle+\langle k\delta_{k}\rangle)}{\langle k^{2}\varphi(k,l)\rangle(\langle l\rangle-\langle l\delta_{l}\rangle)}.\label{23}
\end{equation}

Compare (\ref{23}) with (\ref{4}), we have
\begin{equation}
\bar{\lambda}_{c}=\lambda_{c}+\frac{\langle l^{2}\rangle\langle k\delta_{k}\rangle+\langle kl\rangle\langle l\delta_{l}\rangle}{\langle k^{2}\varphi(k,l)\rangle(\langle l\rangle-\langle l\delta_{l}\rangle)}>\lambda_{c}.\label{24}
\end{equation}

This means that the combined immunization scheme we propose here is indeed effective. Next we will
compare the new immunization scheme with the active immunization scheme,
through (\ref{23}) and (\ref{18}), we have
\begin{align}
\bar{\lambda}_{c}-\tilde{\lambda}_{c}
&=\frac{\langle l^{2}\rangle(\langle k\rangle+\langle k\delta_{k}\rangle)}{\langle k^{2}\varphi(k,l)\rangle(\langle l\rangle-\langle l\delta_{l}\rangle)}-\frac{\langle kl\rangle+\langle l\rangle\langle k\delta_{k,l}^{1}\rangle}{\langle k^{2}\varphi(k,l)\rangle}\notag\\
&=\frac{\langle l^{2}\rangle\langle k\delta_{k}\rangle+\langle kl\rangle\langle l\delta_{l}\rangle+\langle l\rangle\langle l\delta_{l}\rangle\langle k\delta_{k,l}^{1}\rangle-\langle l^{2}\rangle\langle k\delta_{k,l}^{1}\rangle}{\langle k^{2}\varphi(k,l)\rangle(\langle l\rangle-\langle l\delta_{l}\rangle)}.\label{25}
\end{align}

Setting the immunization rate as the same, so we have
\begin{equation}
\overline{\delta}_{l}+\overline{\delta}_{k}=\overline{\delta}_{k,l},\notag
\end{equation}
where $\overline{\delta}_{k,l}$ is defined in Section~\ref{sec-3-2} and $\overline{\delta}_{l},\overline{\delta}_{k}$ are defined in Section~\ref{sec-3-3}. Therefore,
\begin{equation}
 \frac{\langle l\delta_{l}\rangle}{\langle l\rangle}+\frac{\langle k\delta_{k}\rangle}{\langle k\rangle}=\frac{\langle k\delta_{k,l}^{1}\rangle}{\langle k\rangle}.\label{26}
\end{equation}
From (\ref{25}) and (\ref{26}) we obtain that
\begin{align}
\bar{\lambda}_{c}-\tilde{\lambda}_{c}&=\frac{\langle l\delta_{l}\rangle\big(\langle l\rangle\langle k\delta_{k}\rangle+\langle k\rangle\langle l\delta_{l}\rangle\big)}{\langle k^{2}\varphi(k,l)\rangle(\langle l\rangle-\langle l\delta_{l}\rangle)}\notag\\
&>0,\label{27}
\end{align}
as it is obvious that $\langle l\rangle >\langle l\delta_{l}\rangle$.

That is to say, under the same average immunization rate, the combined immunization is more
effective than the active immunization discussed in Section~\ref{sec-3-2}. Fig.~\ref{fig:1}(b)
below illustrates this conclusion.

Considering (\ref{24}) and (\ref{16}), under the same average immunization $\frac{\langle l\delta_{l}\rangle}{\langle l\rangle}+\frac{\langle k\delta_{k}\rangle}{\langle k\rangle}=\langle \delta_{k,l}^{3}\rangle$, we have
\begin{equation}
\bar{\lambda}_{c}=\frac{\langle l^{2}\rangle(\langle k\rangle+\langle k\delta_{k}\rangle)(\langle\varphi(k,l)k\rangle-\langle\varphi(k,l)k\delta_{k,l}^{3}\rangle)}{\langle k^{2}l\varphi(k,l)\rangle(\langle l\rangle-\langle l\delta_{l}\rangle)}\hat{\lambda}_{c}^{3}.\notag
\end{equation}
Note that
\begin{equation}
\langle l^{2}\rangle(\langle k\rangle+\langle k\delta_{k}\rangle)(\langle\varphi(k,l)k\rangle-\langle\varphi(k,l)k\delta_{k,l}^{3}\rangle)>0,\notag
\end{equation}
\begin{equation}
\langle k^{2}l\varphi(k,l)\rangle(\langle l\rangle-\langle l\delta_{l}\rangle)>0.\notag
\end{equation}
So $\bar{\lambda}_{c}=\Lambda\hat{\lambda}_{c}^{3}$, where $\Lambda$ is a positive constant,
which means that the
combined immunization scheme is comparable in effectiveness to the targeted immunization
scheme discussed in Section~\ref{sec-3-1}.

\subsection{Duplex immunization}
\label{sec-3-4}

Infected individuals play a vital part in the early stages of disease transmission, diseases spread
through out-going links of infected individuals to the in-coming links of susceptible individuals.
Therefore the out-degree $l$ is a key character during the early stage of a disease transmission.
And all  immunization strategies~\cite{Pastor2001,Pastor2003,satorras2003,Pastor2002,bai2007}
consider the all nodes as a whole to implement the immunization strategies, so in this section
we proposed an immunization strategy based on a partition of the out-degrees $l$; we divide the
population of all nodes with out-degrees $l$ and in-degrees $k$ into two parts: Nodes with
out-degrees exceeding a positive constant number $L$ is considered as the first
part $T=\{(k,l)|l>L$\}; and $\overline{T}$, the complement of $T$, as the second part. 
In $\overline{T}$, we
use the targeted immunization in Section~\ref{sec-3-1}, and in $\overline{T}$, we use the combined
immunization proposed in Section~\ref{sec-3-2}. It turns out that the effectiveness of these
two immunization strategies' combination is more effective than both of them.

We introduce a constant $\alpha\in[0, 1]$, which indicates the proportion of set $\overline{T}$,
so we have
\begin{center}
  $s_{k,l}(t)+\rho_{k,l}(t)=\begin{cases}
\alpha, &\mbox{if}~~(k,l)\in \overline{T},\\
1-\alpha, &\mbox{if}~~(k,l)\in T.
\end{cases}$
\end{center}

The condition $\alpha= 1$ implies the classic targeted immunization (1) in Section~\ref{sec-3-1}
, while $\alpha= 0$ means the combined immunization proposed in Section~\ref{sec-3-3},
implementing two kinds of immunization strategies together, the model (\ref{1}) becomes
\begin{equation}
\frac{d\rho_{k,l}}{dt}=\begin{cases}
\lambda k(1-\widehat{\delta}_{k})(\alpha-\rho_{k,l}(t))\theta(t)-\rho_{k,l}(t), &\mbox{if}~~(k,l)\in \overline{T},\\
\lambda k(1-\alpha-\rho_{k,l}(t))(1-\overline{\delta}_{l'})\theta(t)-(1+\overline{\delta}_{k'})\rho_{k,l}(t), &\mbox{if}~~(k,l)\in T.\label{28}
\end{cases}
\end{equation}
where
\begin{equation}
  \overline{\delta}_{l'}=\underset{l}{\sum}\frac{lQ(l)}{\langle l \rangle}\delta_{l'},~ \overline{\delta}_{k'}=\underset{k}{\sum}\frac{kP(k)}{\langle k \rangle}\delta_{k'},\notag
\end{equation}
and
\begin{center}
  $\widehat{\delta}_{k}=\begin{cases}
1, ~k>\eta_{1}\\
a', ~k=\eta_{1}~,\\
0, ~k<\eta_{1}
\end{cases}$
    $\delta_{l'}=\begin{cases}
1, ~l>\eta_{2}\\
b', ~l=\eta_{2}~,\\
0, ~l<\eta_{2}
\end{cases}$
 $\delta_{k'}=\begin{cases}
1, ~k>\eta_{3}\\
c', ~k=\eta_{3}~,\\
0, ~k<\eta_{3}
\end{cases}$
\end{center}
where $0<a',b',c'\leq1$, and  $\eta_{2}\ge L$.

By letting $\frac{d\rho_{k,l}}{dt}=0$, than substitute it into (\ref{2}), model (\ref{28})
leads to a self-consistency equation:
\begin{equation}
\Theta=\frac{\lambda\Theta}{\langle l\rangle}\left\{ \underset{k,l}{\sum}\frac{\alpha\lambda\varphi(k,l)p(k,l)k(1-\widehat{\delta}_{k})}{1+\lambda k(1-\widehat{\delta_{k}})\Theta}+\underset{k,l}{\sum}\frac{(1-\alpha)k\varphi(k,l)p(k,l)(1-\overline{\delta}_{l'})}{\lambda k(1-\overline{\delta}_{l'})\Theta+1+\overline{\delta}_{k'}}\right\}\equiv f_{5}(\Theta),\notag
\end{equation}
therefore, we can obtain the epidemic threshold for model (\ref{28}):
\begin{equation}
\check{\lambda}_{c}=\frac{\langle l\rangle}{\alpha\big(\langle\varphi(k,l)k\rangle-\langle\varphi(k,l)k\widehat{\delta}_{k}\rangle\big)}+\frac{\langle l^{2}\rangle(\langle k\rangle+\langle k\delta_{k'}\rangle)}{(1-\alpha)\langle k^{2}\varphi(k,l)\rangle(\langle l\rangle-\langle l\delta_{l'}\rangle)},\label{29}
\end{equation}
which is clearly greater than the epidemic threshold $\lambda_{c}$ obtained in (\ref{4}),
that means the immunization scheme we proposed is indeed effective.

In Section~\ref{sec-3-1}, we know that the TGC is more effective than TGA and TGB, and through
Section~\ref{sec-3-2}, we find the combined immunization is more effective than the active
immunization. 

We now compare the new immunization scheme, the duplex immunization, with the TGC and the
combined immunization to find the optimal one.

Through (\ref{29}) and (\ref{16}), we have
\begin{align}
\check{\lambda}_{c}-\hat{\lambda}_{c}^{3}&=\frac{\langle l\rangle}{\alpha\big(\langle\varphi(k,l)k\rangle-\langle\varphi(k,l)k\widehat{\delta}_{k}\rangle\big)}+\frac{\langle l^{2}\rangle(\langle k\rangle+\langle k\delta_{k'}\rangle)}{(1-\alpha)\langle k^{2}\varphi(k,l)\rangle(\langle l\rangle-\langle l\delta_{l'}\rangle)}-\frac{\langle l\rangle}{\langle\varphi(k,l)k\rangle-\langle\varphi(k,l)k\delta_{k,l}^{3}\rangle}\notag\\
&=\frac{1}{A^{*}}\big( B^{*}+C^{*}-D^{*}\big),\notag
\end{align}
where 
\begin{eqnarray*}
  A^{*} &=& \alpha(1-\alpha)\langle k^{2}\varphi(k,l)\rangle(\langle l\rangle-\langle l\delta_{l'}\rangle)(\langle\varphi(k,l)k\rangle-\langle\varphi(k,l)k\widehat{\delta}_{k}\rangle)(\langle\varphi(k,l)k\rangle-\langle\varphi(k,l)k\delta_{k,l}^{3}\rangle),\\
  B^{*} &=& =\langle l\rangle(1-\alpha)\langle k^{2}\varphi(k,l)\rangle(\langle l\rangle-\langle l\delta_{l'}\rangle)(\langle\varphi(k,l)k\rangle-\langle\varphi(k,l)k\delta_{k,l}^{3}\rangle), \\
  C^{*} &=& \alpha(\langle\varphi(k,l)k\rangle-\langle\varphi(k,l)k\widehat{\delta}_{k}\rangle)\langle l^{2}\rangle(\langle k\rangle+\langle k\delta_{k'}\rangle)(\langle\varphi(k,l)k\rangle-\langle\varphi(k,l)k\delta_{k,l}^{3}\rangle), \\
  D^{*} &=& \langle l\rangle(1-\alpha)\alpha\langle k^{2}\varphi(k,l)\rangle(\langle l\rangle-\langle l\delta_{l'}\rangle)(\langle\varphi(k,l)k\rangle-\langle\varphi(k,l)k\widehat{\delta}_{k}\rangle).
\end{eqnarray*}
It is obvious that these four polynomials: $A^{*}$, $B^{*}$, $C^{*}$ and $D^{*}$ are greater than zero.
For $\overline{T}$, the average immunization rate is $\alpha\langle \widehat{\delta}_{k}\rangle$;
For $T$, the average immunization rate is $(1-\alpha)(\frac{\langle l\delta_{l'}\rangle}{\langle l\rangle}+\frac{\langle k\delta_{k'}\rangle}{\langle k\rangle})$; hence, under the
same average immunization rate for duplex immunization and TGC, we have
\begin{equation}
(1-\alpha)(\frac{\langle l\delta_{l'}\rangle}{\langle l\rangle}+\frac{\langle k\delta_{k'}\rangle}{\langle k\rangle})+\alpha\langle \widehat{\delta}_{k}\rangle=\langle \delta_{k,l}^{3}\rangle.\label{30}
\end{equation}
Note that
\begin{equation}
B^{*}-D^{*}=\langle l\rangle(1-\alpha)\langle k^{2}\varphi(k,l)\rangle(\langle l\rangle-\langle l\delta_{l'}\rangle)\big(\langle\varphi(k,l)k\rangle-\langle\varphi(k,l)k\delta_{k,l}^{3}\rangle-\alpha(\langle\varphi(k,l)k\rangle-\langle\varphi(k,l)k\widehat{\delta}_{k}\rangle)\big).\label{31}
\end{equation}
Combining (\ref{30}) and (\ref{31}), we obtain that when $\alpha$ satisfies the
condition: $\alpha<E^{*}<1$, then
\begin{equation}
\check{\lambda}_{c}>\hat{\lambda}_{c}^{3} ,\label{32}
\end{equation}
where $E^{*}=(\langle\varphi(k,l)k\rangle-\langle\varphi(k,l)k\delta_{k,l}^{3}\rangle)/(\langle\varphi(k,l)k\rangle-\langle\varphi(k,l)k\widehat{\delta}_{k}\rangle)$, which means that the duplex immunization scheme is more effective than the target immunization scheme
discussed in Section~\ref{sec-3-1}.

With the similar analysis method above, we can verify that the duplex immunization scheme is more
effective than the combined immunization scheme discussed in Section~\ref{sec-3-3}.
Fig.~\ref{fig:1}(a) below shows $\check{\lambda}_{c}>\bar{\lambda}_{c}$ very clearly with the
same average immunization rate $\delta=0.12$, and  this result applies to all same $\delta$.

Besides, through \cite{Fu2008,w2012}, we know the targeted
immunization scheme is more effective than proportional 
immunization scheme and acquaintance immunization scheme 
for the same average immunization rate in directed networks. 
Hence, we reach the  following conclusion: \emph{The duplex immunization 
we proposed has the best effectiveness comparing to all other usual 
immunization schemes} with respect to degree distribution in directed scale-free networks.


\subsection{A brief summary}

In the previous section, we have discussed targeted, active, combined and duplex immunization schemes,
and calculated the thresholds for these strategies. By comparing the thresholds for different
immunization strategies, we have conclude that the  epidemic threshold of TGB (see Eq.~(\ref{15}))
is greater than that of TGA (see Eq.~(\ref{14})); Then we proved that the targeted
nodes with both large in-degrees and large out-degrees (see Eq.~(\ref{16})) are 
more worthy to be immunized
in directed networks; We extended the traditional active immunization~\cite{Fu2008} into directed
networks, analyzed its epidemic threshold and compared its effectiveness with targeted immunization
scheme and combined immunization scheme under the same average immunization rate;  The proposed
combined immunization scheme is more effective than active immunization scheme, and it is comparable
to the targeted immunization scheme; And the performance of the duplex immunization scheme is the best
among all usual immunization schemes discussed in this section.


\section{Numerical analysis}
\label{sec-4}

In this section, we present the results of numerical simulations to further illustrate the above
theoretical analysis and show the effectiveness of different immunization schemes. We use the
algorithm of Barab\'{a}si and
Albert~\cite{R2002} to generate a directed scale-free network with $\gamma = 1$ and $ \gamma^{'}=1$.
Here the joint degree distribution is independent, we consider a population of 1000 individuals and
take a unit recovery rate.

\begin{figure}[H]
\begin{minipage}[t]{0.5\textwidth}
\centering
\includegraphics[width=1.0\textwidth,height=.3\textheight]{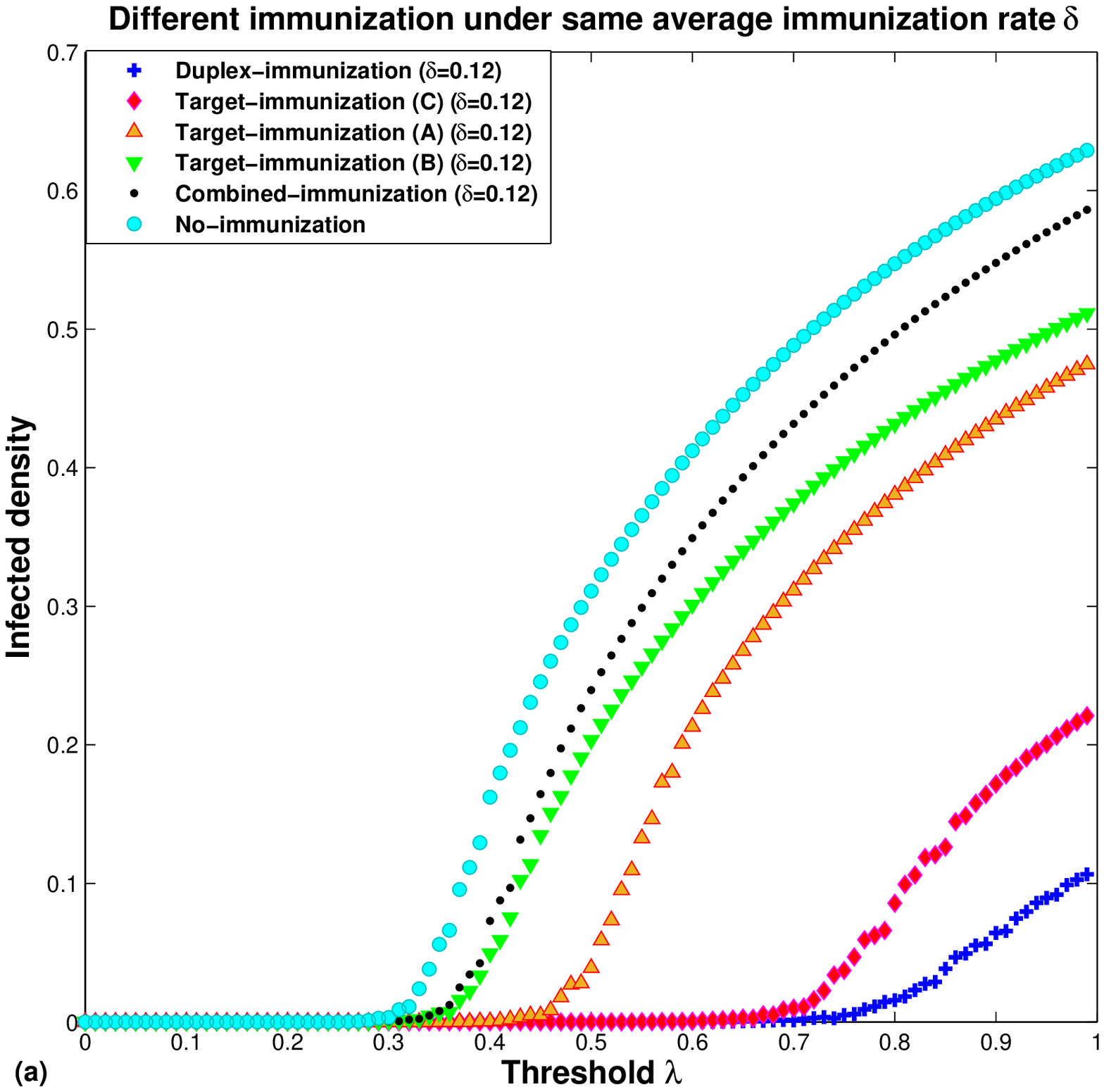}\par{
}
\end{minipage}
\begin{minipage}[t]{0.5\textwidth}
\centering
\includegraphics[width=1.0\textwidth,height=.3\textheight]{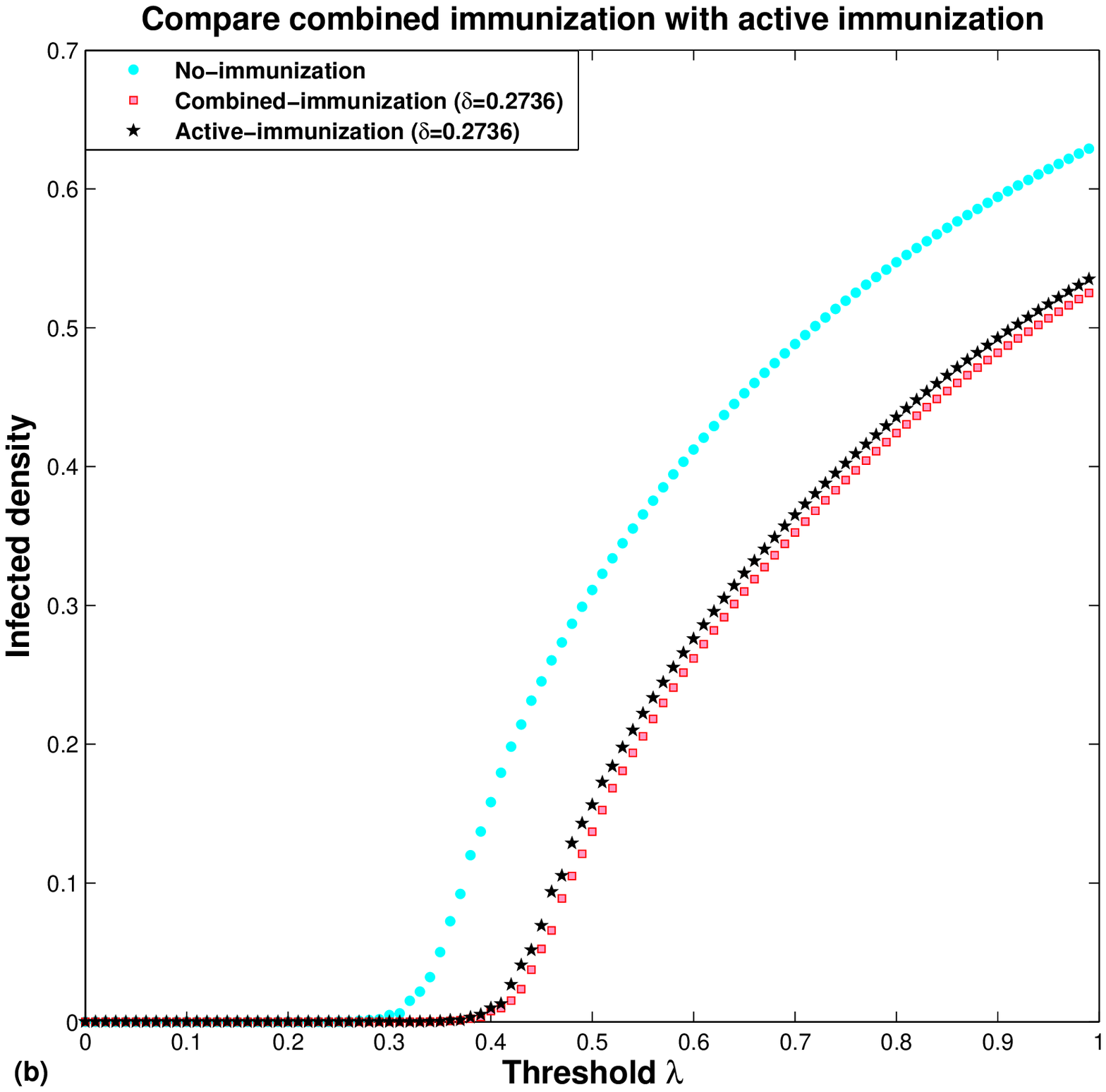}\par{
}
\end{minipage}\caption{\footnotesize (Color online) Comparison of the effectiveness of different
immunization schemes under the same average immunization rate $\delta$. (a) shows the thresholds among
those  five immunization schemes under the average immunization rate $\delta=0.12$, the threshold $\lambda_c=0.25$ for no immunization, the threshold  $\bar{\lambda}_{c}=0.27$ for combined
immunization, the threshold $\hat{\lambda}_{c}^{2}=0.29$ for TGB,  the threshold 
 $\hat{\lambda}_{c}^{1}=0.4$ for TGA, the threshold $\hat{\lambda}_{c}^{3}=0.6$ for TGC,  the
threshold $\check{\lambda}_{c}=0.7$ for duplex immunization. (b) shows that the combined
immunization scheme is  more effective than the active scheme for the same average immunization
rate $\delta=0.2736$, the threshold $\tilde{\lambda}_{c}=0.32$ for active immunization
and $ \bar{\lambda}_{c}=0.35$ for combined immunization.}
 \label{fig:1}
\end{figure}

In Fig.~\ref{fig:1}(a), we repeated the simulation above when the immunization
 schemes-targeted (1, 2, 3), combined and duplex-are implemented. We set the same average immunization
rate for these five immunization schemes to  better compare their effectiveness with each other, and
we implement a no-immunization curve for all immunization schemes to illustrate that all immunization
schemes are effective comparing to the case without any immunization. The average
out-degree ($\langle k\rangle$) and average in-degree ($\langle l\rangle$) for the generated network
is 3 and 5, respectively, and $k_{max}=82, l_{max}=100$. Here we set the infectivity as a constant $\varphi(k,l)=2$.
For TGA we choose $\pi_{1}=7$ and $a=1$, for TGB
we choose $\pi_{2}=4$ and $b=1$, and for TGC we choose $\pi_{3}=4$,  $\pi_{4}=2$
and $c=1$. We can see in Fig.~\ref{fig:1}(a), under the same average immunization
rate $\delta=0.12$, $\hat{\lambda}_{c}^{3}>\hat{\lambda}_{c}^{1}>\hat{\lambda}_{c}^{2}$, which means the
performance of TGC is better than TGA and TGB. And for the targeted immunizations on a directed
scale-free network, to immunize nodes with large out-degrees is more efficient than to immunize
nodes with large in-degrees.

\begin{figure}[H]
\begin{minipage}[t]{0.5\textwidth}
\centering
\includegraphics[width=0.8\textwidth,height=.21\textheight]{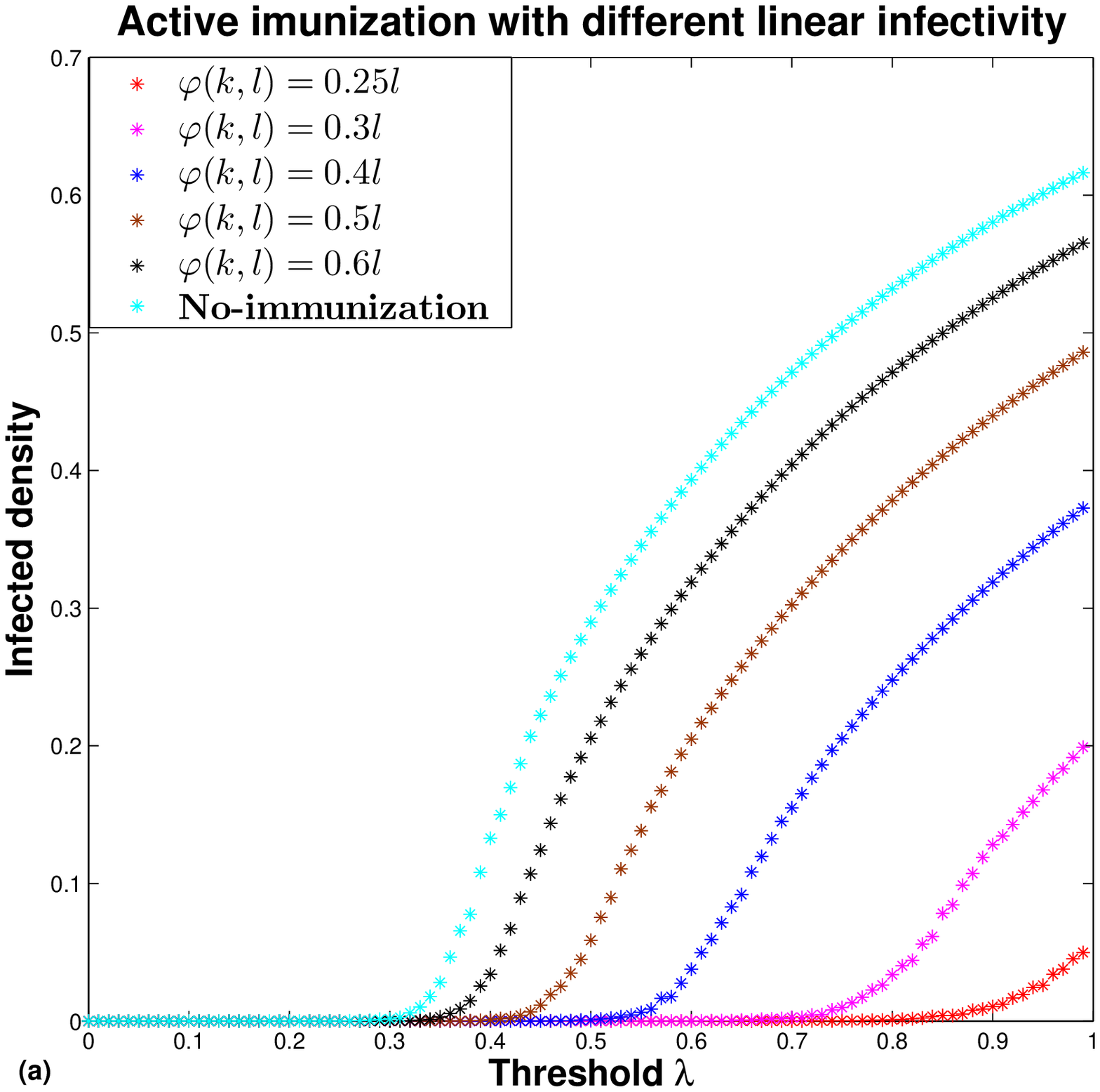}\par{
}
\end{minipage}
\begin{minipage}[t]{0.5\textwidth}
\centering
\includegraphics[width=0.8\textwidth,height=.21\textheight]{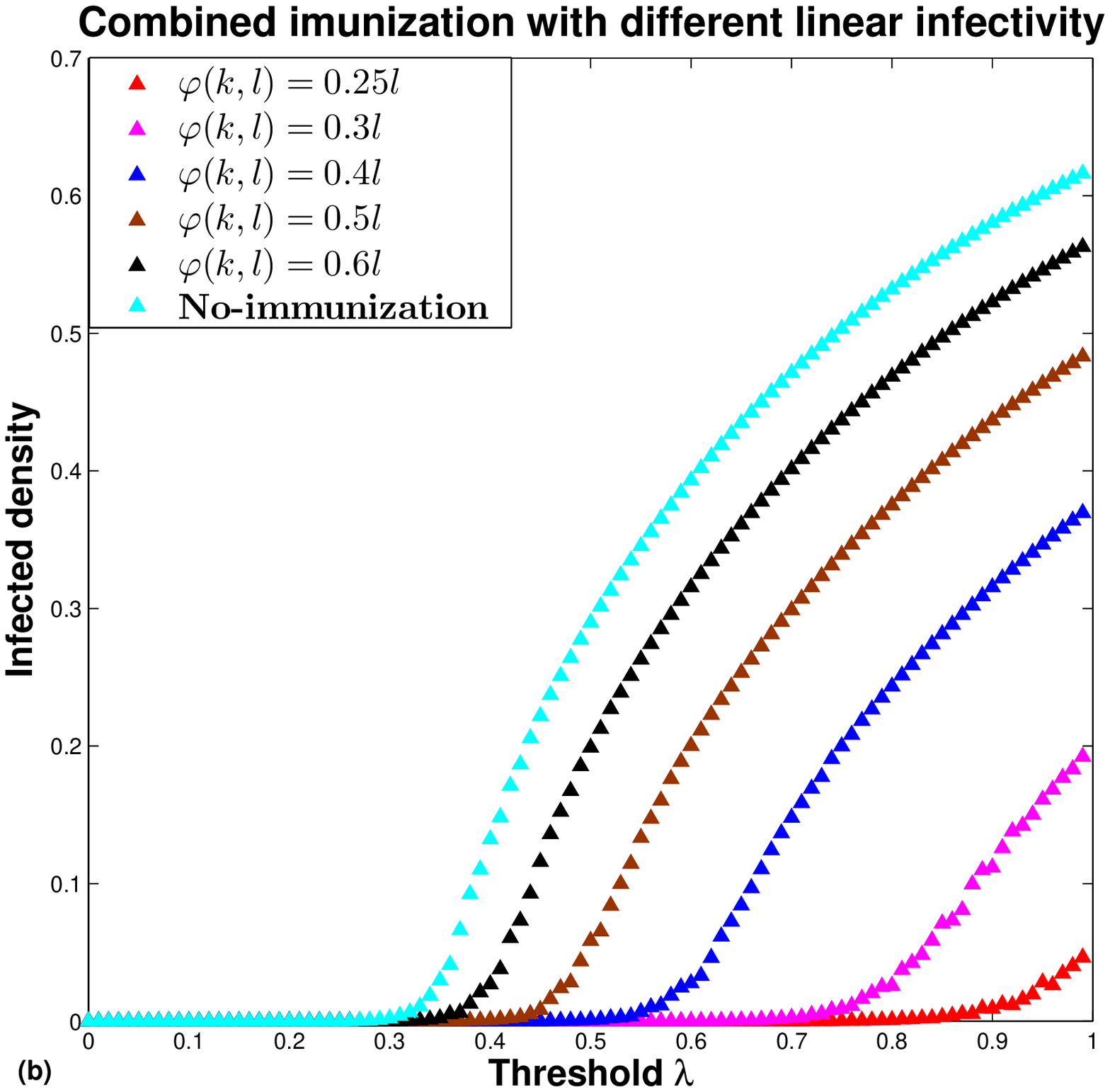}\par{
}
\end{minipage}
\begin{minipage}[t]{0.5\textwidth}
\centering
\includegraphics[width=0.8\textwidth,height=.21\textheight]{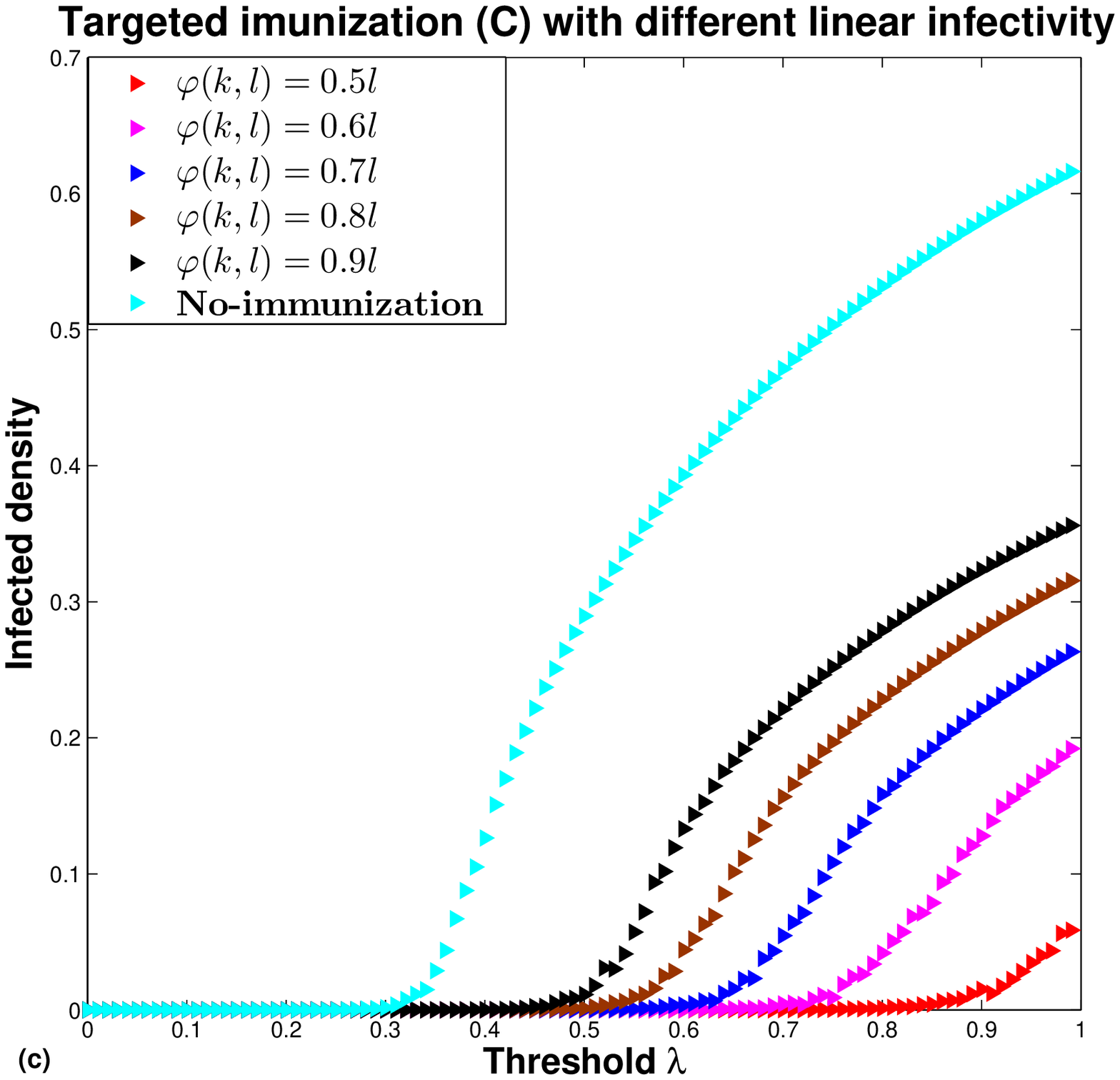}\par{
}
\end{minipage}
\begin{minipage}[t]{0.5\textwidth}
\centering
\includegraphics[width=0.8\textwidth,height=.21\textheight]{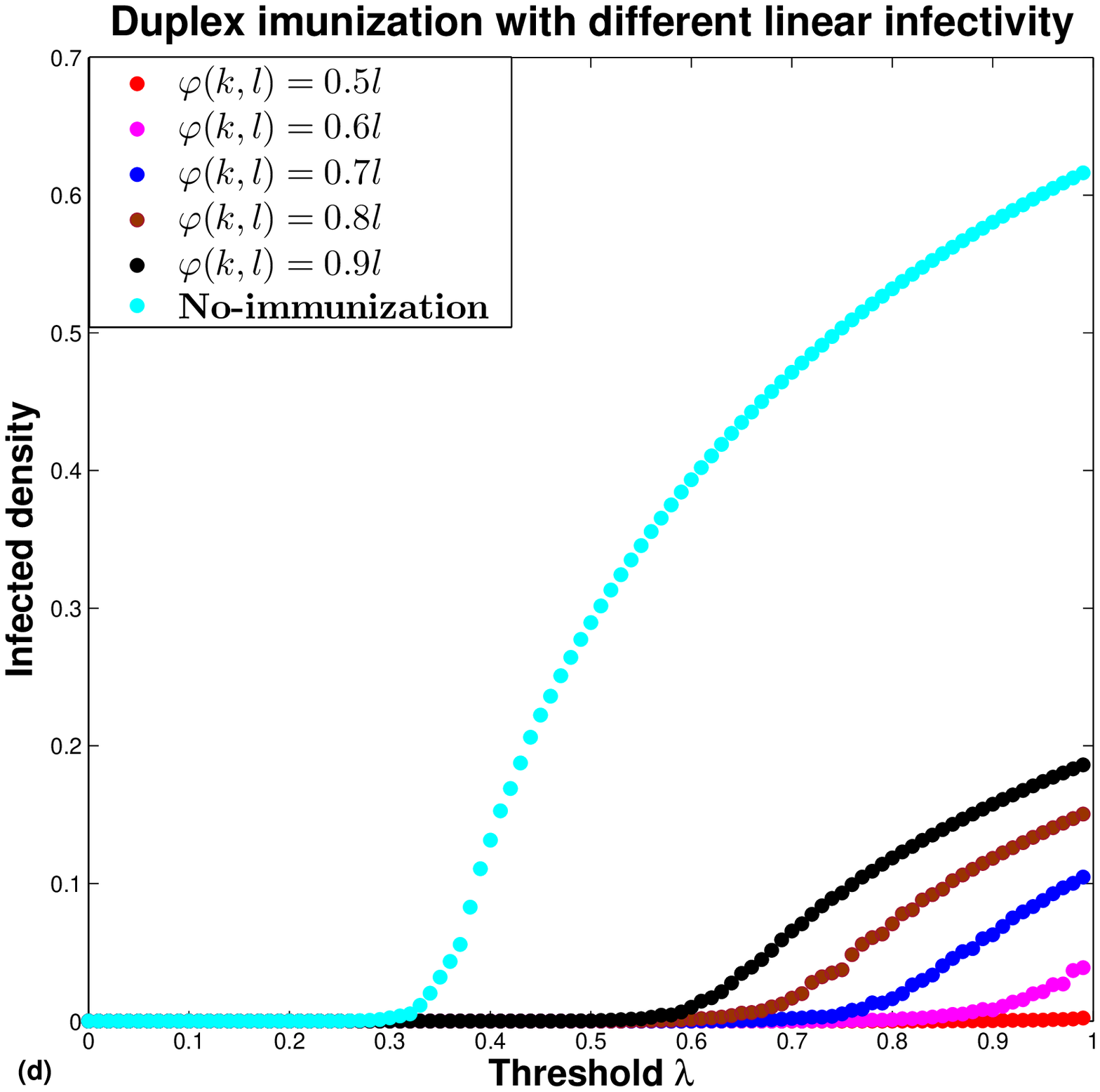}\par{
}
\end{minipage}\caption{\footnotesize(Color online) Different linear infectivity $\varphi(k,l)$'s effects on four immunization schemes. We take linear infectivity $\varphi(k,l)=al$,  in (a) and (b), $a=0.25, 0.3, 0.4, 0.5, 0.6$;  in (c) and (d), $a=0.5, 0.6, 0.7, 0.8, 0.9$.}
 \label{fig:2}
\end{figure}

\begin{figure}[H]
\begin{minipage}[t]{0.5\textwidth}
\centering
\includegraphics[width=0.8\textwidth,height=.21\textheight]{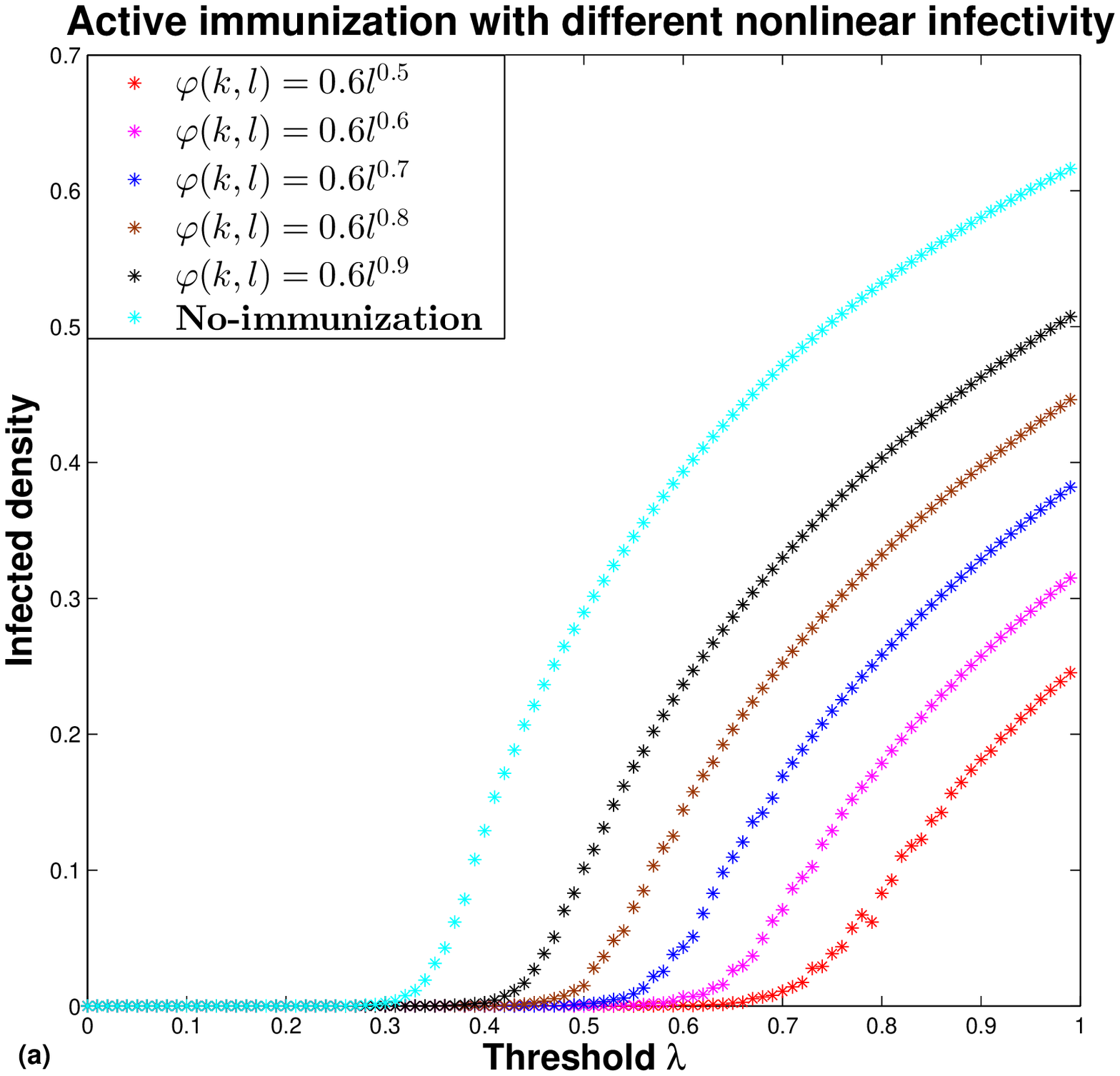}\par{
}
\end{minipage}
\begin{minipage}[t]{0.5\textwidth}
\centering
\includegraphics[width=0.8\textwidth,height=.21\textheight]{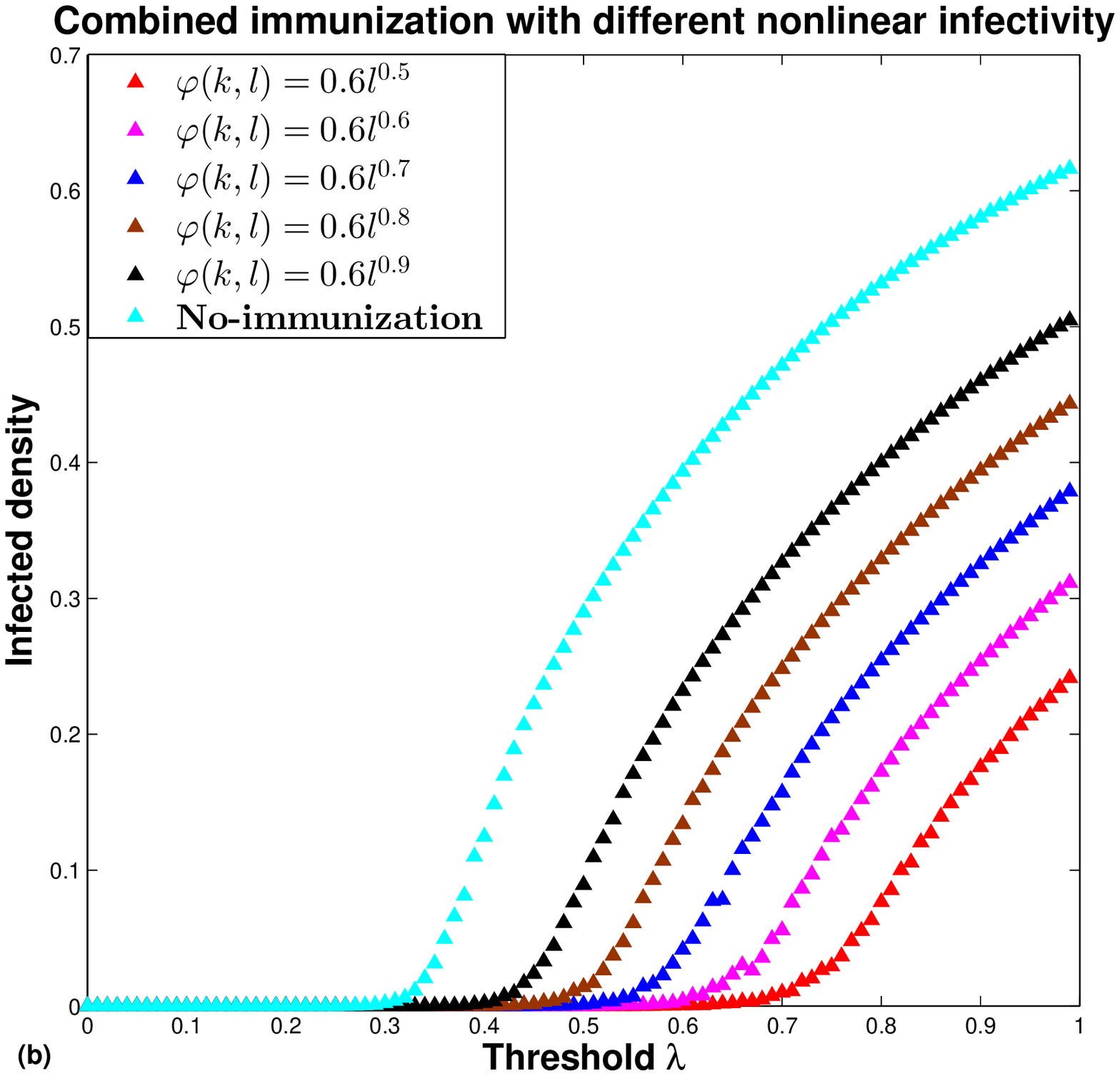}\par{
}
\end{minipage}
\begin{minipage}[t]{0.5\textwidth}
\centering
\includegraphics[width=0.8\textwidth,height=.21\textheight]{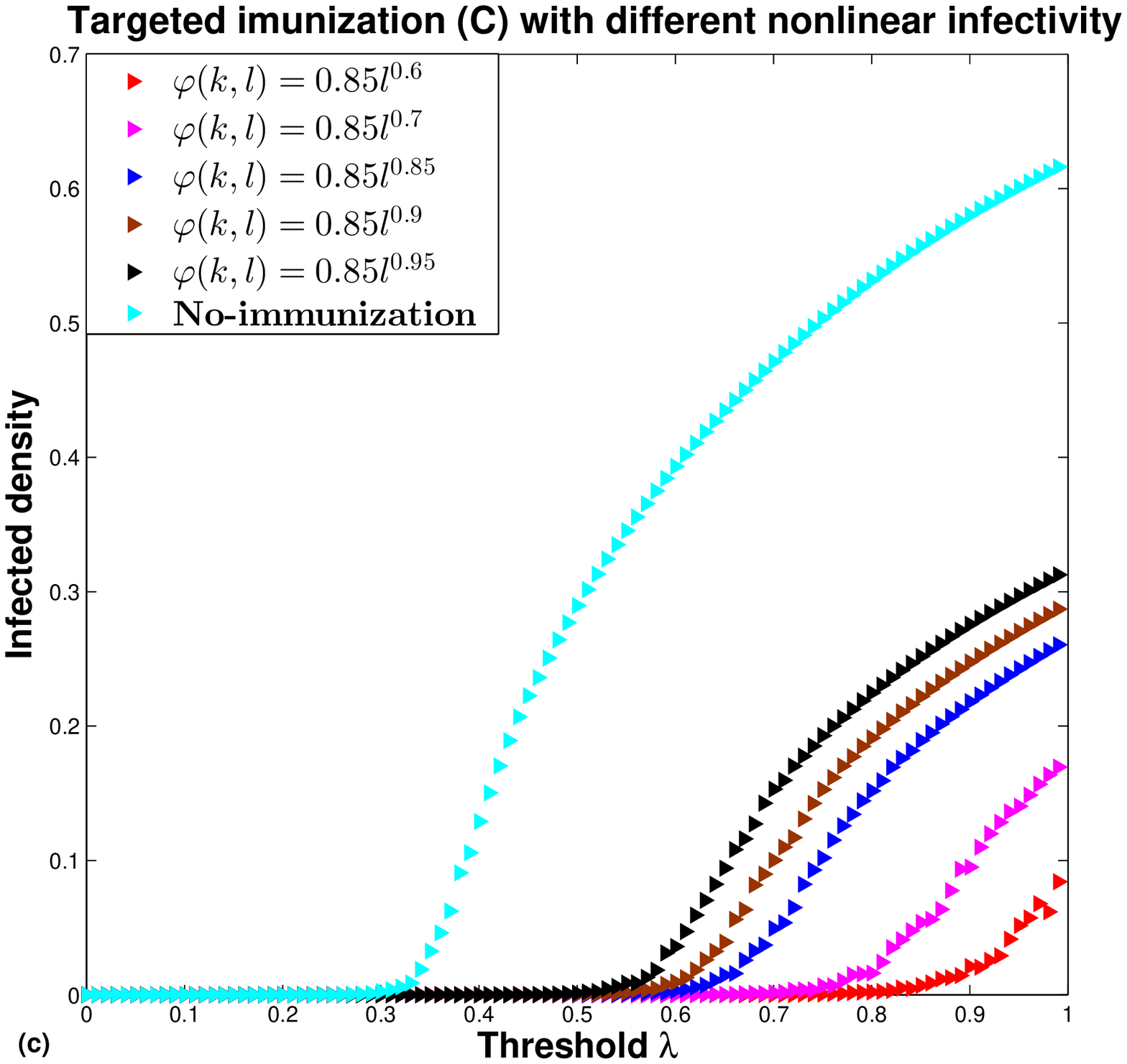}\par{
}
\end{minipage}
\begin{minipage}[t]{0.5\textwidth}
\centering
\includegraphics[width=0.8\textwidth,height=.21\textheight]{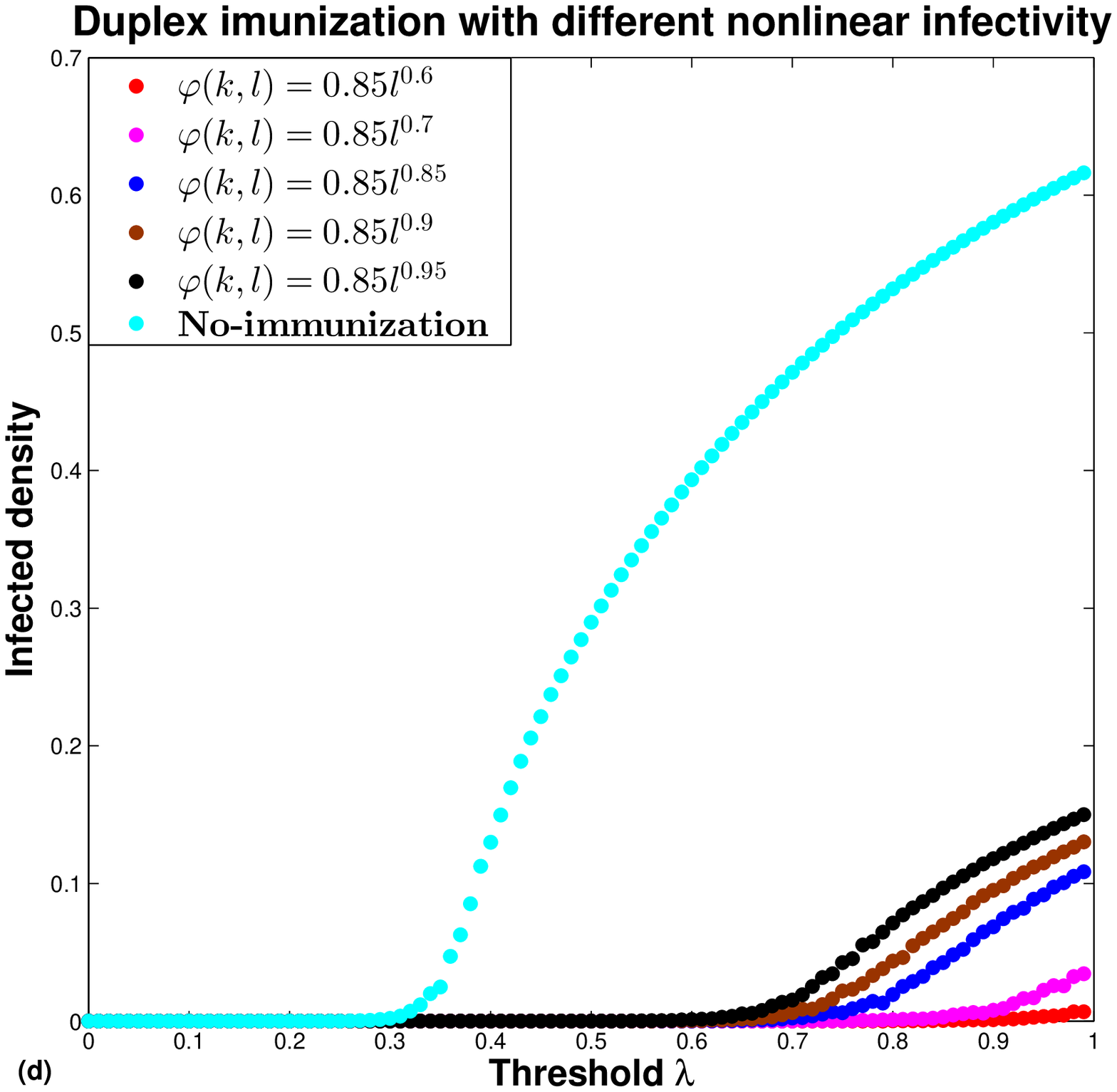}\par{
}
\end{minipage}\caption{\footnotesize(Color online) Different nonlinear infectivity $\varphi(k,l)$'s
effects on four immunization schemes. We take nonlinear infectivity $\varphi(k,l)=al^{\alpha}$,
in (a) and (b), $a=0.6$, $\alpha=0.5, 0.6, 0.7, 0.8, 0.9$;  in (c)
and (d), $a=0.85$, $\alpha=0.6, 0.7, 0.85,  0.9, 0.95$.}
 \label{fig:3}
\end{figure}

Besides, we can obtain the epidemic threshold for duplex immunization $\check{\lambda}_{c}=0.7$
and $\hat{\lambda}_{c}^{3}=0.6$ of TGC, which verifies the conclusion in Section~\ref{sec-3-4}, and means
that the duplex immunization we proposed is more effective than the targeted immunization 
discussed in~\ref{sec-3-1} for the same average immunization rate.
Here we take $T=\{(k,l)|l>2$\}, $\eta_{1}=17$, $a'=1$, $\eta_{2}=18$, $b'=1$, $\eta_{3}=10$, $c'=1$,
and $\alpha$ can be calculated as $0.6188$. When infectivity takes linear or nonlinear
function, in Figs.~\ref{fig:2}(c-d) and Figs.~\ref{fig:3}(c-d), those conclusions are
still valid.

In Fig.~\ref{fig:1}(b), we compare the thresholds among no immunization, active and combined
immunization schemes; we set the same average immunization rate $\delta=0.2736$ for those two
immunization schemes. Here $\pi_{1}=9$ for active immunization scheme,
and $\kappa_{1}=15, \kappa_{2}=13$ for combined immunization scheme. So we can illustrate the
conclusion in Section~\ref{sec-3-3} (see Eq.~\ref{27}), which means that the combined immunization
scheme proposed in Section~\ref{sec-3-3} is more effective than the active immunization scheme
discussed in Section~\ref{sec-3-2} for the same average immunization rate.

In Fig.~\ref{fig:2}, we choose a linear infectivity $\varphi(k,l)=al$ and set the
same $\delta=0.12$. For active and combined immunization schemes, we set $a=0.25, 0.3, 0.4, 0.5, 0.6$,
and for targeted immunization(c) and duplex immunization, we set $a=0.5, 0.6, 0.7, 0.8, 0.9$.
Still, in Figs.~\ref{fig:2}(a)-(d), we take the same average immunization rate $\delta=0.12$.
The Figs.~\ref{fig:2}(a)-(d) clearly show that with an increasing $a$, the thresholds
of those four immunization schemes are increasing at the same time; on the other hand, with
different linear infectivity, the duplex immunization is still more effective than targeted
immunization, and the combined immunization is still more effective than active immunization.

In Fig.~\ref{fig:3}, we choose a nonlinear infectivity $\varphi(k,l)=al^{\alpha}$. For active
and combined immunization, we set $a=0.6$, $\alpha=0.5, 0.6, 0.7, 0.8, 09$ and for targeted
immunization(c) and duplex immunization, we set $a=0.85$, $\alpha=0.5, 0.6, 0.7, 0.8, 09$ .
We choose $\delta=0.12$ in Figs.~\ref{fig:3}(a)-(d), which shows the similar property when
infectivity is linear: when $\alpha$ increases, threshold increases. And by different
nonlinear infectivity, the thresholds changes faster than linear infectivity. Besides,
with different nonlinear infectivity for same immunization rate, the duplex immunization
is more effective than targeted immunization and the combined immunization is more effective
than active immunization.

In addition, in Fig.~\ref{fig:1}, we present the contrast among different immunization strategies
for the same immunization rate and validate that the duplex immunization is more effective than
targeted immunization; the performance of combined immunization is better than active immunization.
In Figs.~\ref{fig:2}-\ref{fig:3}, we use different linear and nonlinear infectivities $\varphi(k,l)$
on active, combined, targeted and duplex immunization schemes, it is shown that with higher
infectivity, the epidemic threshold is dramatically reduced; besides, the results of
comparison between those immunization schemes are still valid with different linear and
nonlinear infectivities.

\section{Conclusions and discussions}
\label{sec-5}

In this paper, different immunization strategies for SIS models in directed scale-free networks
with different infectivities are studied, and we calculate the epidemic thresholds for different 
immunization schemes, and obtained the following results:

Firstly, the epidemic threshold $\lambda_c$ takes a positive value if $\varphi(k,l)=al^{\alpha}$
and $\alpha<\gamma'$ in a finite network with sufficiently large size; besides, when $\varphi(k,l)=\frac{al^{\alpha}}{1+bl^{\alpha}}\cdot\frac{ck^{\beta}}{1+dk^{\beta}}$, $\lambda_c$
is always positive.

Secondly, for the targeted immunization in directed networks, we prove that immunizing nodes with
large out-degrees
are more effective than immunizing nodes with large in-degrees when targeted immunization is
implemented; on the other hand, we demonstrate that the nodes with both large in-degrees and large
out-degrees are more worthy to be immunized during target immunization process than nodes
with only large in-degrees or large out-degrees.

Thirdly, the duplex immunization we proposed has the best effectiveness comparing to all other usual immunization schemes (e.g., proportional immunization, acquaintance immunization, 
targeted immunization, active immunization, and combined immunization)
for the same average immunization rate.
Besides, the performance of the combined immunization we proposed on disease control is better
than active immunization.

Finally, from realistic viewpoints, weighted networks and degree-correlated networks are
more reasonable for epidemic immunization, and we expect that our work may be extended 
into these and even multiplex and interconnected networks in our future research.

\subsection*{Acknowledgements}

This work was jointly supported by the NSFC under grants 11572181 and 11331009.


\end{document}